\documentclass[prb,longbibliography,twocolumn]{revtex4-1}
\usepackage{geometry}
\usepackage{amsmath}
\usepackage{amssymb}
\usepackage{graphicx}
\usepackage{hyperref}
\usepackage{braket}
\usepackage{bm}
\usepackage{xcolor}

\newcommand{\be}{\begin{equation}}
\newcommand{\ee}{\end{equation}}
\newcommand{\bk}{{{\bf{k}}}}
\newcommand{\bK}{{{\bf{K}}}}

\newcommand{\br}{{{\bf{r}}}}

\newcommand{\bR}{{{\bf{R}}}}
\newcommand{\bq}{{\bf{q}}}

\newcommand{\bea}{\begin{eqnarray}}
\newcommand{\eea}{\end{eqnarray}}
\newcommand{\beal}{\begin{align}}
\newcommand{\eeal}{\end{align}}
\newcommand{\ra}{\rangle}
\newcommand{\la}{\langle}

\newcommand{\upa}{\uparrow}
\newcommand{\dna}{\downarrow}

\newcommand{\dg}{{\dagger}}
\newcommand{\pdg}{{\phantom\dagger}}

\newcommand{\vf}{v_{\rm f}}
\newcommand{\vd}{v_{\Delta}}
\newcommand{\mh}{\mathfrak{h}}
\newcommand{\vvf}{\vec{v}_{\rm f}}
\newcommand{\vvd}{\vec{v}_{\Delta}}
\def\veps{\varepsilon}

\begin{document}

\title{Pseudo-Landau levels of Bogoliubov quasiparticles in strained nodal superconductors}
\author{Geremia Massarelli}
\affiliation{Department of Physics, University of Toronto, Toronto,
  Ontario M5S 1A7, Canada}
  \author{Gideon Wachtel}
\affiliation{Department of Physics, University of Toronto, Toronto,
  Ontario M5S 1A7, Canada}
\author{John Y. T. Wei} 
\affiliation{Department of Physics, University of Toronto, Toronto,
  Ontario M5S 1A7, Canada}
  \author{Arun Paramekanti} 
\affiliation{Department of Physics, University of Toronto, Toronto,
  Ontario M5S 1A7, Canada}
\begin{abstract}
Motivated by theory and experiments on strain induced pseudo-Landau levels (LLs) of Dirac fermions in graphene and topological materials, we consider its extension for Bogoliubov quasiparticles (QPs) in a nodal superconductor (SC). We show, using an effective low energy description and numerical lattice calculations for a $d$-wave SC, that a spatial variation of the electronic hopping amplitude or a spatially varying $s$-wave pairing component can act as a pseudo-magnetic field for the Bogoliubov QPs, leading to the formation of pseudo-LLs. We propose realizations of this phenomenon in the cuprate SCs, via strain engineering in films or nanowires, or $s$-wave proximity coupling in the vicinity of a nematic instability, and discuss its signatures in tunneling experiments.
\end{abstract}

\maketitle

\section{Introduction}

The ability to tune electronic properties with strain in a wide range of quantum materials has led to the emerging area of `straintronics' \cite{StraintronicsReview_2016}. Strain has been shown to be an important knob in graphene, topological materials, and oxide electronics, allowing one to tune band dispersion and topology \cite{Neto2009Electronic, Guinea_NPhys2010,Vozmediano2010Gauge,Levy2010Strain,Manoharan_Nature2012,Madhavan_NNano2015,Ramesh_PRB2016,Okamoto_PRB2016,SYang_NanoLett2016}, and to control magnetism \cite{Ramesh_PRB2013,YJKim_PRL2014} and ferroelectricity \cite{Choi_Science2004} in thin films. Uniaxial strain has also been used to shed light on fundamental questions in correlated materials, from searching for chiral $p_x \pm i p_y$ pairing in Sr$_2$RuO$_4$ \cite{Hicks2014Strong}, to understanding nematicity in pnictide superconductors \cite{Kuo_Science2016} and in the `hidden order' state of URu$_2$Si$_2$ \cite{Riggs_NComm2015}.

In graphene, a two-dimensional (2D) electronic membrane \cite{Kim_EPL2008}, strain modifies the wavefunction overlap between neighboring orbitals and causes a momentum space displacement of the massless Dirac point in the dispersion, thus simulating the effect of a vector potential~\cite{Neto2009Electronic,Vozmediano2010Gauge,Naumis_RPP2017}. A spatial variation of the strain in graphene nanobubbles and `artificial graphene' leads to colossal  pseudo-magnetic fields of up to $\sim \!\! 300$T, and a pseudo-Landau level (pseudo-LL) spectrum~\cite{Guinea_NPhys2010, Levy2010Strain, Manoharan_Nature2012}.
Strain also induces a deformation potential which acts as a `scalar gauge potential'; the corresponding in-plane electric fields can lead to a breakdown of the pseudo-LLs \cite{Lukose_PRL2007,Sanjuan_PRB2014,Castro_2016,Naumis_RPP2017}. There have been theoretical studies of Josephson coupling through pseudo-LLs  \cite{Covaci_PRB2011,Uchoa_PRB2015}, and interaction effects which can lead to exotic correlated states \cite{Ghaemi2012Fractional,Uchoa_PRL2013}.
Strain effects have also been generalized to 3D Dirac and Weyl semimetals \cite{Cortijo_PRL2015,Cortijo_PRB2016,Garate_2016,Liu2017Quantum}, Kitaev spin liquids \cite{Rachel_PRL2016}, and atoms in optical lattices \cite{ColdAtomGauge_RMP2011,Pekker2015}.

In light of these developments, we address in this paper the important question of how these phenomena manifest themselves in superconducting phases of matter. Specifically, we consider the possibility of engineering time-reversal invariant pseudo-gauge fields for Bogoliubov quasiparticle (QP) excitations of nodal superconductors (SCs).
Our key observation is that the QP Dirac nodes of the SC will shift in momentum space under the modification of the single-particle dispersion or the form of the pairing gap. Thus, spatial variations of the dispersion or the pairing term can mimic a spatially varying gauge field. Using an effective low energy theory for 2D $d$-wave SCs
%\cite{Nersesyan1994Disorder}, 
as well as a numerical lattice model study, we show that this induces pseudo-LLs of Bogoliubov QPs and discuss its signatures in the spatially resolved
tunneling density of states (TDOS).

Our work highlights two key differences between strained nodal SCs and materials such as graphene or Dirac-Weyl semimetals. (i) Unlike electrons, Bogoliubov QPs do not have a well-defined electrical charge and do not couple directly to external orbital magnetic fields. Thus, strain engineering provides a unique window to explore LL physics of Bogoliubov QPs. (ii) We show that strain variations in a $d$-wave SC with time-reversal symmetry cannot induce a pseudo-`scalar potential' for Bogoliubov QPs. This is unlike the impact of the deformation 
potential for graphene. In this regard, pseudo-LLs of Bogoliubov QPs are more robust and are `symmetry protected'.

We suggest two routes to realizing this physics in the cuprate SCs: via strain engineering in thin films and nanowires, or via edge effects or $s$-wave proximity coupling in the vicinity of an isotropic to nematic SC quantum phase transition (QPT) \cite{Lawler_PRB2008}. Our study sheds light on how inhomogeneous strain can reorganize the low energy spectrum of nodal SCs. 

\section{Effective low-energy theory}

\begin{figure}[tb]
\centering
\makebox[0.1\textwidth]{\includegraphics[width=0.5\textwidth]{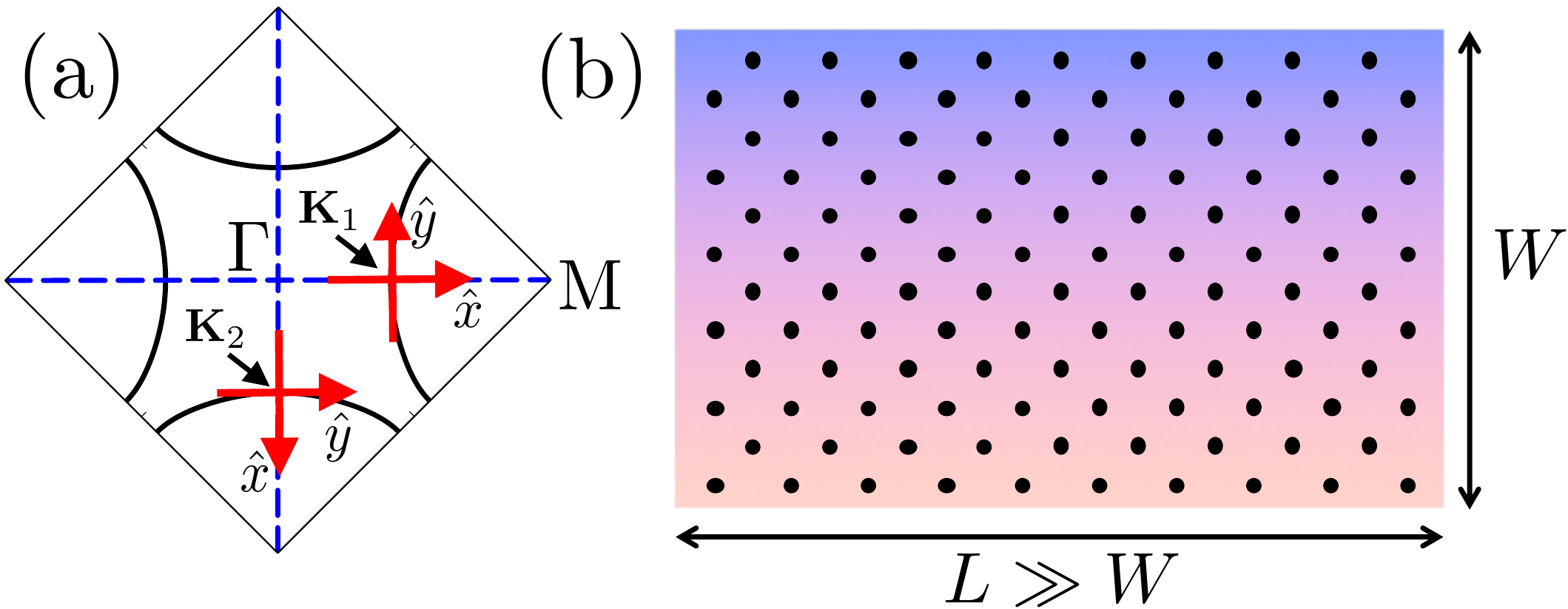}}
\caption{(Color online) (a) Rotated Brillouin zone for the square lattice showing schematic Fermi surface (solid, black) for optimal hole-doped cuprate 
SCs. Quasiparticle Dirac nodes are located at $\pm \bK_1$ and $\pm \bK_2$, and we show local coordinate axes used in our low-energy theory. 
(b) Strip geometry (not to scale) used in the numerics with width $W$ and length $L\gg W$. Shading gradient illustrates spatial variation in the pairing or hopping amplitude across the strip.}
\label{fig:BZ_strip}
\end{figure}

The low energy excitations of a uniform 2D $d$-wave SC on a square lattice reside near the two pairs of gap nodes 
$\bK_{\pm 1}\! \equiv \! \pm (K,K)$
and $\bK_{\pm 2} \!\equiv \! \pm (K,-K)$ as in Fig.~\ref{fig:BZ_strip}(a).
We combine the slowly varying fermion fields near the node pairs into Nambu spinors 
$\Psi^\dg_{\ell\alpha}(\br) \equiv (\psi^\dg_{\ell\alpha}(\br),\epsilon_{\alpha\nu} \psi^\pdg_{-\ell\nu}(\br))$, where $\alpha,\nu$ are
spin labels ($\upa$ or $\dna$), and $\ell=1,2$ labels the nodes $\bK_{1,2}$. The low energy
excitations of a nodal SC are
described by the effective Dirac Hamiltonian $H_0=\sum_{\ell,\alpha} \int d^2\br \Psi^\dg_{\ell\alpha}(\br) \mh^{(\ell)}_0 \Psi^\pdg_{\ell\alpha}(\br)$, with
\bea
\mh^{(\ell)}_0 &=& - i \sigma^z \vvf^{(\ell)} \!\cdot\! \vec \nabla - i \sigma^x \vvd^{(\ell)}\! \cdot \! \vec \nabla
\eea
where $\vvf^{(\ell)},\vvd^{(\ell)}$ denote the Fermi velocity and the gap velocity (respectively, normal and tangential to the
Fermi surface), and $\sigma^{x,z}$ are Pauli matrices.
Diagonalizing $H_0$ in momentum space
leads to the massless Dirac dispersion $E_\ell(\bk) = (\vf^2 k^2_x + \vd^2 k^2_y)^{1/2}$ where $(k_x,k_y)$ denotes the deviation in momentum
from $\bK_\ell$ (with local coordinate axes as shown in Fig.~\ref{fig:BZ_strip}(a)), and the Dirac cone anisotropy is set by $\vf/\vd$.

We next turn to the effect of time-reversal invariant slow spatial variations in the hopping and pairing amplitudes of this nodal SC, which adds to the microscopic lattice Hamiltonian terms of the form
\bea
\! \delta H_1 \!\!&=&\! - \frac{1}{2} \sum_{\bR,\eta,\alpha} \delta t^\pdg_\eta(\bR) (c^\dg_{\bR,\alpha} c^\pdg_{\bR+\eta,\alpha} \!+\! {\rm h.c.}) \\
\! \delta H_2 \!\!&=&\!\!  \frac{1}{8} \! \sum_{\bR,\eta} \!\! \delta\Delta^\pdg_\eta(\bR) (c^\dg_{\bR\upa} c^\dg_{\bR+\eta,\dna} \!-\! c^\dg_{\bR\dna} 
c^\dg_{\bR+\eta,\upa} \!+\! {\rm h.c.})
\eea
where $\eta$ denotes the set of neighbors of site $\bR$ and `h.c.' stands for Hermitian conjugate. A low energy expansion of the fermion fields leads to the modified Hamiltonian
\bea
\mh^{(\ell)} \!=\! \begin{pmatrix} - i \vf \partial_x + f_\ell(\br) & - i s_\ell \vd \partial_y + g_\ell(\br) \\ - i  s_\ell \vd \partial_y + g_\ell(\br) & i \vf \partial_x - f_\ell(\br) \end{pmatrix}
\eea
where $s_\ell=(-1)^\ell$, with
\bea
f_\ell(\br) \!&=&\! - \sum_\eta \delta t_\eta(\br) \cos(\bK_\ell\cdot\eta) \\ %\equiv f_{\rm e}(\br) + s_n f_{\rm o}(\br)
g_\ell(\br) \!&=&\! \frac{1}{4} \sum_\eta \delta \Delta_\eta(\br) \cos(\bK_\ell\cdot\eta), %\equiv g_{\rm e}(\br) \!+\! s_n g_{\rm o}(\br)
\eea
and we have implicitly assumed that we have rotated $\br$ into the local coordinate axes for node $\ell$. 
{Note that a conventional deformation potential or spatially varying chemical potential may also be included in $f_{\ell}(\br)$ in Eq.~4}.
We can recast this Hamiltonian as
\bea
\!\!\!\! \mh^{(\ell)} \!=\! \vf \sigma^z  (-i \partial_x \!+\! {\cal A}^{(\ell)}_{x}\!(\br)) \!+\! s_\ell \vd \sigma^x 
(-i \partial_y \!+\! {\cal A}^{(\ell)}_y\!(\br))
\eea
where we have defined the `vector potential' $\vec{\cal A}^{(\ell)}$ via $\vf {\cal A}^{(\ell)}_{x}\!(\br) \equiv f_\ell(\br)$ and $\vd {\cal A}^{(\ell)}_{y}\!(\br) \equiv s_\ell g_\ell(\br)$. 
Thus slow spatial modulations of parameters in a nodal superconductor will lead to an effective low energy theory of Dirac quasiparticles coupled to a 
spatially varying `vector potential'.

{The issue of whether additional gauge potentials (e.g. a `scalar gauge potential' which minimally couples to time-derivatives rather than space derivatives) can arise in a strained SC amounts to asking if any other Pauli matrix components are permitted in $\mh^{(\ell)}$. To address this, we note that terms proportional to the identity matrix will act as a valley-odd chemical potential, while a component proportional to $\sigma^y$ will correspond to complex pairing. Both terms are forbidden by time-reversal and spin-rotation symmetries in a $d$-wave SC, and thus cannot destabilize the pseudo-LLs; in this sense, the pseudo-LLs may be regarded as `symmetry protected' (see Appendix A for details). 
The key point is that slow modulations of the parameters of a nodal superconductor will leave the nodal quasiparticle excitations pinned to zero energy but 
can displace it in momentum space.
Thus, $d$-wave Bogoliubov QPs, unlike electrons in graphene, do not experience an inhomogeneous `scalar' gauge potential~\cite{Naumis_RPP2017,Castro_2016}. However, breaking time-reversal symmetry, for instance with a supercurrent, will lead to a Doppler shift for the QPs~\cite{deGennesBook}, shifting the energy of the nodal
excitations, which thus provides an analog of a `scalar potential'.}

\section{Pseudo-Landau levels}

We next turn to the spectrum of $\mh^{(\ell)}(\br)$ for two illustrative cases, with $\vec{\cal A}$ induced by variations in the pairing gap or hopping amplitude, to show the emergence of pseudo-LLs. We then supplement the continuum theory with numerical results on a lattice realization.

\subsection{Pseudo-LLs from gap variations} 
Let us impose an additional extended $s$-wave pairing with a uniform gradient along the $[1,1]$ direction, which translates to $\delta \Delta_{+x}(\br)\!=\! \delta \Delta_{+y} (\br) \!=\! (x_a/a_0 + x_b/a_0 + 1/2) \Delta_s$. 
Here, $(x_a,x_b)$ refer to (global) coordinates corresponding to the $a$ and $b$ crystal axes, and $a_0$ is the lattice constant. 
Using this, we find $f_\ell(\br)\!=\! 0$, while, in the local coordinates at $\ell\!=\!1,2$, we have $g_{1}(\br)\!=\! \beta \vd x$ and $g_{2}(\br)\!=\!\beta \vd y$, with $\beta \equiv \sqrt{2}  \frac{\Delta_s}{\vd a_0} \cos K $. 

For node pair $\ell \!=\! 2$, this leads to $\vec {\cal A}^{(2)}\!=\!(0, \beta y)$, which yields $\vec {\cal B}^{(2)}\! =\! 0$. In this case, the energy spectrum is unaffected by the modulation, while the wavefunctions are obtained by a gauge rotation as ${\rm e}^{- \frac{i}{2} \beta y^2} \Psi^{(2)}(\br)$, where $\Psi^{(2)}(\br)$ is the Nambu spinor wavefunction of the uniform $d$-wave SC for node pair $\ell=2$.

For node pair $\ell\!=\!1$, we arrive at $\vec {\cal A}^{(1)}\!=\!(0, -\beta x)$, i.e., the Landau gauge for a pseudo-magnetic field $\vec {\cal B}^{(1)} \!=\! - \beta \hat{z}$. Setting the Nambu wavefunction $\Psi^{(1)}(\br) \!=\! {\rm e}^{i k y} \Phi^{(1)}(x)$, we get (see Appendix B)
\be
\!\!\! \left[\! - i \vf  \sigma^z \partial_x \!+ \! \beta \vd \sigma^x (x \!-\! \frac{k}{\beta}) \! \right] \Phi^{(1)}(x) = E \Phi^{(1)}(x).
\ee
Defining
$\ket{\upa} \!=\! \frac{1}{\sqrt{2}} (1, i~{\rm sgn}\beta)^T$ and $\ket{\dna} \!=\! \frac{1}{\sqrt{2}} (1, -i~{\rm sgn}\beta)^T$,
we find a
zero energy eigenstate
$\ket{\Phi_{k0}}= \ket{0}_{k} \ket{\dna}$
and nonzero energy eigenstates
\bea
\ket{\Phi_{kn\pm}} &=& \frac{1}{\sqrt{2}} \big( \ket{n-1}_{k} \ket{\upa} \pm i \ket{n}_{k} \ket{\dna}\big),
\eea
where the subscript $\pm$ denotes states with energies
$\pm \sqrt{2|\beta| \vd\vf n}$ (with integer $n \geq 1$).
Here, $\ket{n}_{k}$ is the $n^{\rm th}$
eigenstate of a harmonic oscillator centered at $k/\beta$, with a mean square width $\la x^2 \ra = (n+1/2) \frac{\vf}{|\beta|\vd}$. We
confirm these findings below within a lattice model of a $d$-wave superconducting strip.

\subsection{Pseudo-LLs from hopping variations} 

Next, let us consider a uniform spatial gradient in the hopping along the $[1,1]$ direction, given by $\delta t_{+x}(\br) = \delta t_{+y}(\br) = - (x_a/a_0 + x_b/a_0 +1/2) t_s$, where $t_s$ sets the scale of the hopping distortion.
This results in $g_\ell(\br)\! =\! 0$ and, in local coordinates, $f_{1}(\br)\!=\! \beta \vf x$ and $f_{2}(\br)\!=\! \beta \vf y$, where $\beta \equiv 4 \sqrt{2} \frac{t_s}{\vf a_0} \cos K$. 
This, in turn, leads to $\vec {\cal A}^{(1)}\!=\!( \beta x,0)$, which corresponds to zero pseudo-magnetic field, while $\vec {\cal A}^{(2)}\!=\!(\beta y,0)$ yields a pseudo-magnetic field $\vec {\cal B}^{(2)} = -\beta \hat{z}$, which supports pseudo-LL energies identical to the case with gap variation for the same choice of $\beta$ (see Appendix C).
A similar pseudo-vector potential can also be realized by a spatially varying nematic distortion of the second-neighbor hopping, with $\delta t_{+x+y}(\br) = - (x_a/a_0 + x_b/a_0 +1) t_s$ and $\delta t_{+x-y}(\br) = (x_a/a_0 + x_b/a_0) t_s$, which yields $\vec {\cal B}^{(1)}\!=\!0$ and $\vec {\cal B}^{(2)}\!=-\beta \hat{z}$, with $\beta \equiv 4 \sqrt{2} \frac{t_s}{\vf a_0} \sin^2 K$. We note that while these examples are `gauge equivalent' to the earlier gap variation case, their physical realizations are distinct since we are changing the hopping rather than the gap, thus directly controlling the `vector potential'.

\section{Lattice model results}

\begin{figure}[t]
\centering
\makebox[0.1\textwidth]{\includegraphics[width=0.5\textwidth]{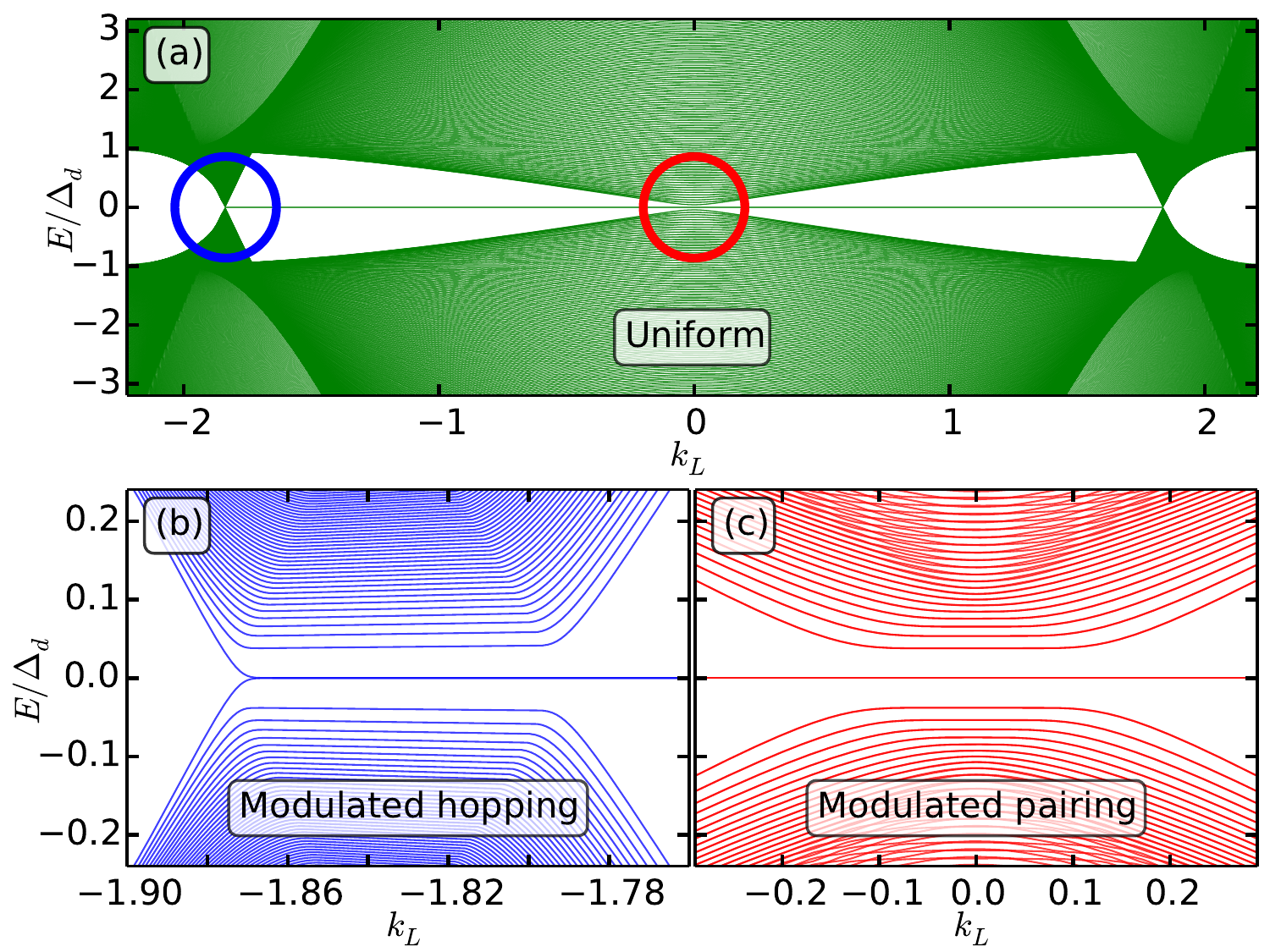}}
\caption{(Color online) 
(a) Spectrum of uniform $d$-wave SC on a $(1,1)$-edged strip versus momentum $k_L$ along the $L$-direction, showing Dirac nodes and zero energy ABSs. Circles indicate regions shown in the next two panels.
(b) Formation of flat pseudo-Landau levels near the outer Dirac nodes due to uniform hopping-amplitude gradient in the [1,1] direction; shown here is the near-node region indicated in (a). 
(c) Similar to (b) but with extended $s$-wave pairing gradient, which induces pseudo-LLs near the central Dirac node indicated in panel (a).}
\label{fig:GradSpectra}
\end{figure}

To check the validity of the low-energy linearized Dirac theory, we numerically diagonalized the full lattice Bogoliubov-deGennes (BdG) Hamiltonian using a strip geometry with $(1,1)$ edges (see Fig.~\ref{fig:BZ_strip}(b)). The strip width is $W$; the transverse direction, along which periodic boundary conditions were used, has length $L \gg W$. Analogous 
results for the $(1,0)$-edged strip are presented in Appendix F.
We pick a nearest neighbor hopping amplitude $t\!=\! 1$, next-neighbor hopping $t'\!=\! -0.25t$, electron filling $\bar{n}\!=\! 0.85$, and a $d$-wave gap $\Delta_d\!=\! 0.25 t$, such that
$\vf/\vd \! \approx\! 13$; these parameters are chosen so as to be representative of the hole-doped cuprate SCs.

Fig.~\ref{fig:GradSpectra}(a) shows the spectrum of the $(1,1)$-edged strip as a function of the momentum $k_L$ along the
long direction $L$, in the absence of any imposed spatial variation for $W\!=\! 500 \sqrt{2} a_0$. 
The spectrum exhibits $d$-wave Dirac nodes projected onto the Brillouin zone of the strip; the velocity anisotropy $\vf/\vd \!\gg\! 1$ is evident in the dispersion slopes of the outer versus inner nodes. In addition, we find zero energy Andreev bound states (ABSs)
expected for a $d$-wave SC in this geometry \cite{Tanaka_RPP2000,Tsuei_RMP2000,Lofwander_SST2001,Deutscher_RMP2005}. 

Fig.~\ref{fig:GradSpectra}(b) shows the spectrum with a nonzero gradient in the hopping amplitude across the strip width, which leads to a pseudo-LL spectrum at the outer Dirac nodes; we have chosen to plot the spectrum near the Dirac node indicated by the circle in Fig.~\ref{fig:GradSpectra}(a), for strip width $W\!=\! 500 \sqrt{2} a_0$ and a maximum change $\delta t \!\sim \! 0.1 t$ at the edge.
Fig.~\ref{fig:GradSpectra}(c) shows the effect of an extended $s$-wave pairing gradient along the strip width,
which leads to pseudo-LL formation at the central Dirac node. Here, we have chosen
$W\!=\! 2000 \sqrt{2} a_0$ and a maximum $s$-wave gap $\Delta_s \!\sim\! 0.4 \Delta_d$ at the edge. The low energy 
spectra in Fig.~\ref{fig:GradSpectra}(b) and (c) are in quantitative agreement with our analytical results. The spectrum
for the $(1,0)$-edged strip (see Appendix F) displays similar strain induced pseudo-LLs; the key
difference is in the absence of ABSs for the unstrained $d$-wave SC in this geometry.

\section{Experimental signature of pseudo-LLs} 

\begin{figure}[tb]
\centering
\makebox[0.1\textwidth]{ \includegraphics[width=0.51\textwidth]{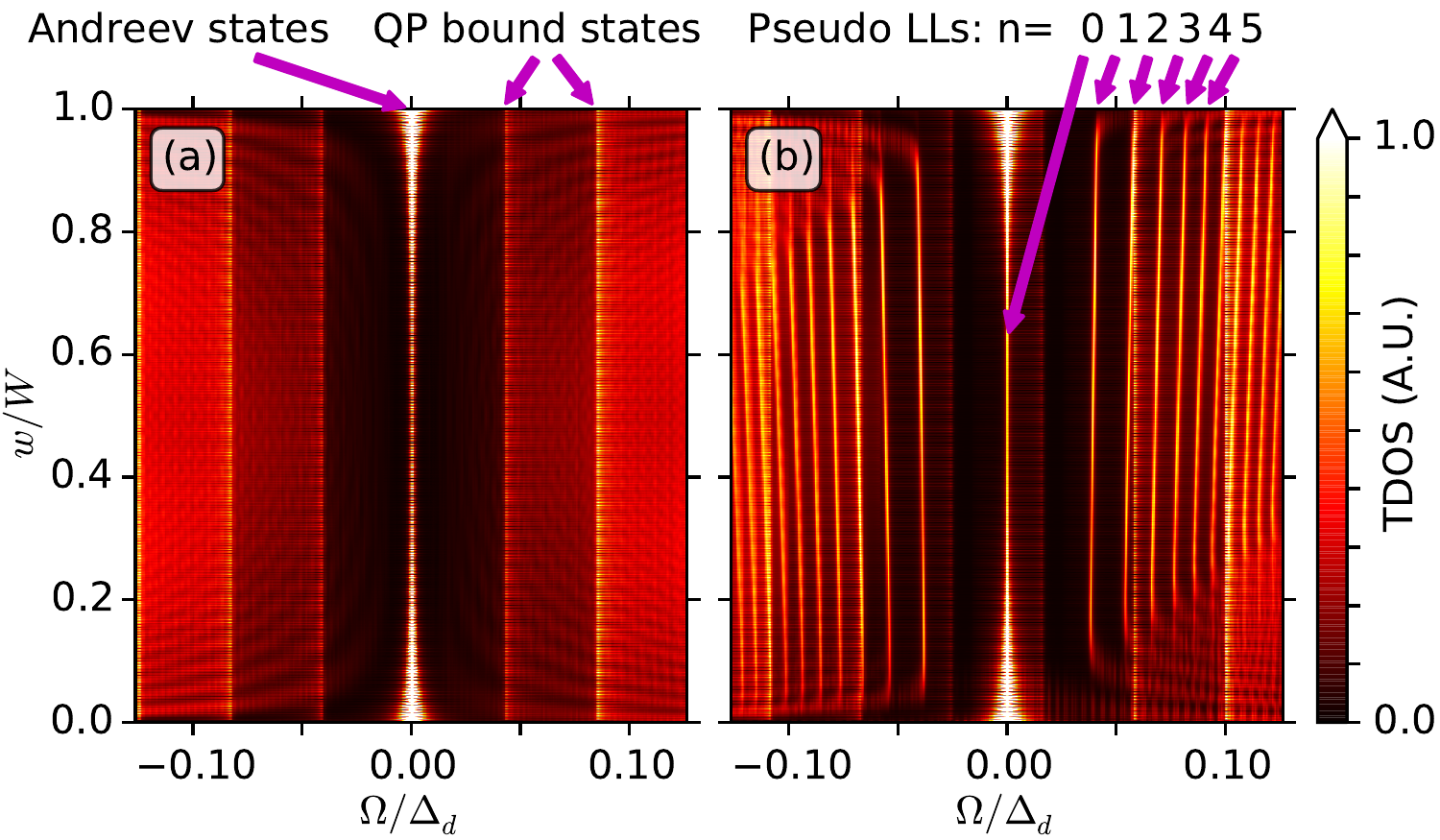} }
\caption{(Color online) Low energy TDOS versus energy $\Omega/\Delta_d$ (scaled to the $d$-wave gap), from diagonalization of BdG 
Hamiltonian in the strip geometry, plotted across scaled strip width $0 \!<\! w/W \!<\! 1$. (a) Uniform $d$-wave SC, showing 
ABSs near zero energy localized near $w/W\!=\!0,1$, and QP bound state TDOS exhibiting rapid spatial oscillations.
(b) Hopping gradient case showing extra pseudo-LL peaks.}
\label{fig:LDOSPanels}
\end{figure}

As in the case of strained graphene, scanning tunneling spectroscopy (STS) experiments which probe the TDOS may provide the most direct route to observing the QP pseudo-LLs. For weak pseudo-magnetic fields, the peaks in density of states due to pseudo-LLs may be visible in microwave spectroscopy. Below, we first provide analytical expressions for the bulk TDOS expected
within our continuum low energy theory. We then present numerical results on the lattice model (see Fig.~\ref{fig:LDOSPanels})
which goes beyond the continuum
theory by incorporating the effects of quantum confinement of the Bogoliubov QPs to the strip, as well as the impact of ABSs
at the edges.

In tunneling experiments, the TDOS in the continuum theory will have two contributions in the bulk.
At nodes where the vector potential acts as pure gauge, it will only induce a phase shift for the fermion operators, leading to a TDOS
contribution identical to a uniform $d$-wave SC. At nodes where the QPs sense a pseudo-magnetic field, there will be discrete
pseudo-LLs. These lead to a total TDOS  (details in Appendix D)
\bea
\!\! N(\Omega) &\approx& \frac{|\Omega|}{\pi \vf \vd} \!+\! \frac{|\beta|}{\pi} \sum_{n} \delta(\Omega-\lambda_n)
\eea
where $n=0,\pm 1, \pm 2, \ldots$, and $\lambda_n = \sqrt{2\beta \vf\vd |n|} {\rm sgn}(n)$.

We have also computed the TDOS numerically for the lattice model in the above strip geometry.
Confinement to the strip then leads to QP subbands with minima at discrete energies $\sim p \pi\vd/W$ and $\sim p \pi\vf/W$ for nodes $\bK_1, \bK_2$ respectively ($p=$ nonzero integer), as well as ABSs at the strip edges.
%For pseudo-LL gap at node $\bK_1$ to dominate over the confinement induced energy scale arising from this node, 
%we need $W \! \gg \! (1/r) (\vf/\vd)$ where $r$ is the fractional change in the pairing or hopping going from the center to the
%edge of the strip. 
As seen from Fig.~\ref{fig:LDOSPanels}, the TDOS for the strip exhibits three key features. 
(i) Without or with a gradient in the hopping amplitude, we see the zero energy peaks in the TDOS at the top and bottom edges reflecting the presence of ABSs; the spectral weight from these ABSs weakly leaks into the bulk.  As shown in Appendix F, the ABSs and their contribution to the TDOS is absent for a (1,0)-edged strip.
(ii) In the bulk (i.e., away from the edges), one set of indicated peaks exhibits rapid spatial oscillation of the TDOS across the strip width. 
These peaks arise when the energy $\Omega$ crosses the minimum $\Omega_0^s$ (at $k_L\!=\!0$) of each subband $s$ in the spectrum, leading to a $\sim\! 1/\sqrt{\Omega \!-\! \Omega_0^s}$ divergence in the TDOS. 
These QP bound states (see Appendix E) arise due to internode scattering $\bK_{1} \leftrightarrow - \bK_{1}$. 
There are additional weaker features with longer-length-scale spatial variations arising from intranode scattering at $\pm \bK_2$. Both contributions are present even in the absence of a gradient; see Fig.~\ref{fig:LDOSPanels}(a). 
(iii) Finally, the hopping gradient induces an extra set of indicated pseudo-LL peaks seen in Fig.~\ref{fig:LDOSPanels}(b) where the TDOS is nearly constant across the strip. The spatial dependence of the TDOS distinguishes the pseudo-LL peaks from QP bound states.

\section{Experimental realizations}

\begin{figure}[t]
\centering
\makebox[0.1\textwidth]{\includegraphics[width=0.35\textwidth]{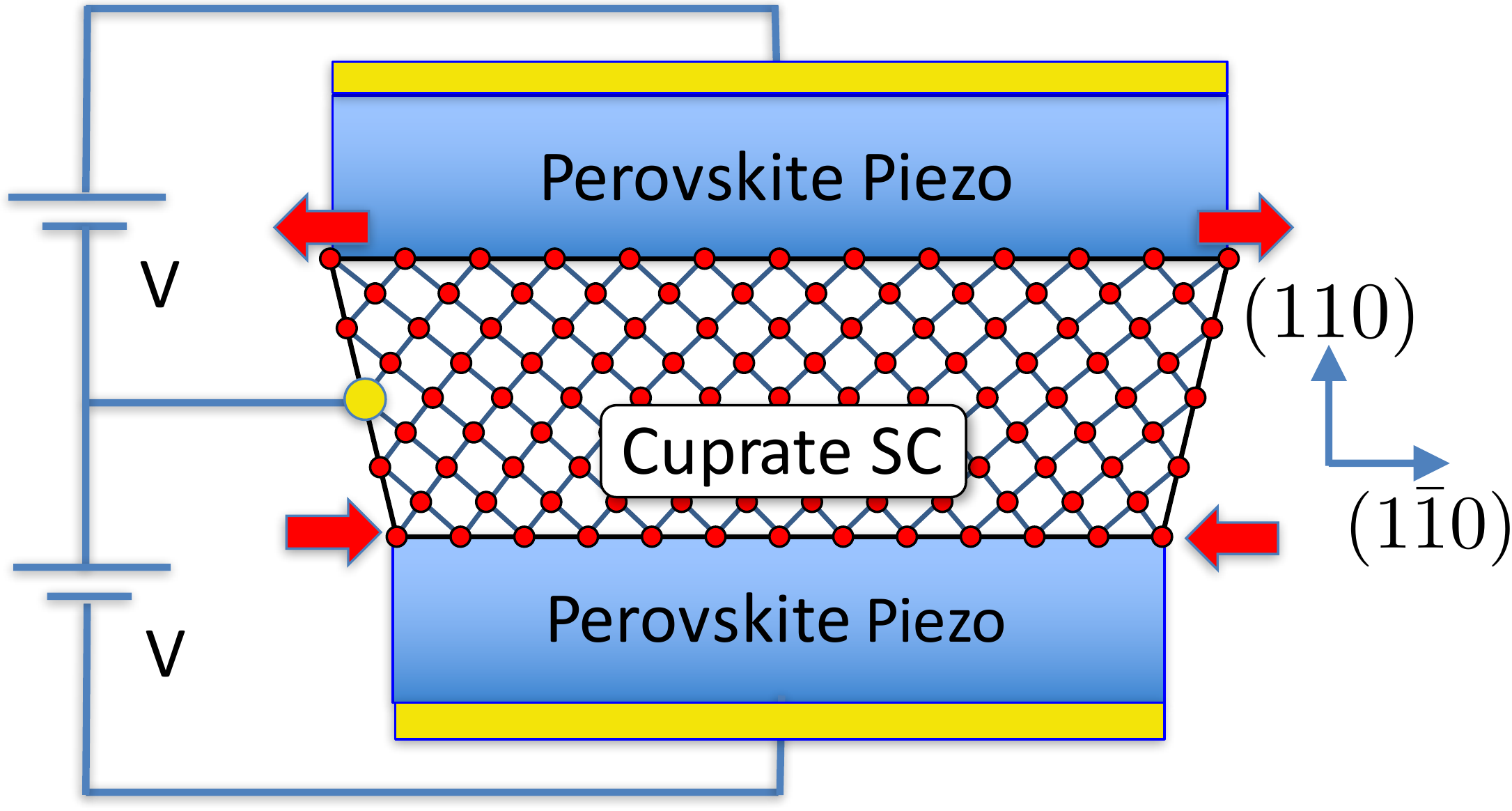}}
\caption{(Color online) Trilayer heterostructure with cuprate SC thin film epitaxially sandwiched between two piezoelectric perovskite films along the $(110)$ surface.  
An inhomogeneous strain can be induced in the cuprate layer by asymmetrically polarizing the two piezo layers.  Metallic outer gates (yellow regions) 
are used to apply the piezo voltages, with the cuprate layer serving as the common inner gate.  For typical values of piezo constant ($d_{31}\! \sim \! 50$-$275$pm/V) 
and dielectric breakdown field ($\sim \! 25$MV/m) for piezoelectric perovskites \cite{hooker1998properties,Trolier-McKinstry2004} 
such as Pb(Zr$_x$Ti$_{1-x}$)O$_3$, we estimate that lattice strains $\sim \! 0.1$-$1 \% $  can be induced in the cuprate layer.}
\label{fig:Piezo}
\end{figure}

\subsection{Strained nanowires or films}

One route to tuning the spatial variation of the electron hopping and pairing amplitudes discussed above is to strain a cuprate thin film or
nanowire. Unlike graphene, which has a 
simple single-particle description of its electronic bands, it is necessary here to include electron interactions in order to study
the microscopic impact of strain on the $d$-wave SC.
The cuprates may be modelled by a $tJ$ Hamiltonian, 
$H_{tJ} \!=\! - g_t \sum_{i,j,\sigma} t^\pdg_{ij} c^\dg_{i\sigma} c^\pdg_{j\sigma} \!+\! g_J J \sum_{\la ij \ra} \vec S_i \cdot \vec S_j$, 
with {\it bare} nearest and next-neighbor hoppings
$t_0$ and $t'_0 \!\approx\! -0.3t_0$ respectively, and nearest-neighbor spin exchange $J \!=\! 4 t_0^2/U \!\approx 0.3 t_0$.
We set $t_0=450$meV which leads to $J = 135$meV.
The coefficients $g_t, g_J$ represent renormalization factors
that crudely account for strong correlation effects. Motivated by slave-boson \cite{Kotliar_PRB1988} and renormalized mean field theory
calculations \cite{RMFT_1988,RMFT_Review2004}, we pick
$g_t\!=\! 2 p/(1\!+\! p)$ and $g_J\!=1$, where $p$ is the hole doping (see Appendix G for details).
Such a mean field approach captures a variety of experimental observations on the $d$-wave cuprate SCs;
we therefore view it as a useful tool to estimate the pseudo-LL gap.

Here, we consider the effects of inhomogeneous strain that can be induced using a piezoelectric thin-film  heterostructure 
schematically depicted and discussed in Fig.~\ref{fig:Piezo}.
Such piezo-induced strain will lead to a gradient in the hopping $\delta t_0(\br)$ as well as a change in the superexchange interaction 
$\delta J(\br) \! \approx  \! (8 t_0/U) \delta t_0(\br)$ across the strip. This induces a gradient in the effective hopping and pairing amplitude in the
BdG equation. Raman scattering studies of La$_2$CuO$_4$ under hydrostatic pressure \cite{Aronson_PRB1991} indicate that
a $\mp 0.5\%$ change in the lattice constant leads to $\delta J/J \approx \! \pm 5\%$, indirectly implying a change in the bare hopping amplitude
$\delta t_0/t_0 \! \approx\! \pm 2.5\%$  in the underlying $tJ$ model. 
A self-consistent solution to the mean field equations
in the SC state at a hole doping
$p\!=\! 0.15$ shows that such a uniform change leads to a $\approx \! \pm 7\%$ change in the $d$-wave pairing gap and $\approx \! \pm 3\%$ change in the
renormalized hopping. A gradient in the $d$-wave gap does not induce any pseudo-LLs; however, the hopping gradient can in fact induce
pseudo-LLs as discussed above.
For a $(110)$-edged film of thickness  $\sim \! 700 a_0$, or a nanowire of similar width ($\approx 270$nm) which is experimentally
realizable \cite{Bonetti_PRL2004} and
similar to the strip geometry explored here, we estimate that a hopping gradient with a realistic $0.5$-$1\%$ maximal strain across the sample will generate a first excited pseudo-LL at $E_1 \! \sim \! 1$meV; this can be probed by $c$-axis tunneling. A fully self-consistent inhomogeneous BdG study of this physics 
is challenging due to the large system sizes involved; we defer this to future work.

\subsection{Proximity to nematic order}

A different route to realizing pseudo-LLs is to note that the onset of nematic order in a tetragonal $d$-wave SC spontaneously breaks the $C_4$ point group symmetry 
and will induce an extended $s$-wave component to the pair field \cite{Lawler_PRB2008}. There is evidence that the cuprates are
proximate to such a QPT \cite{Hinkov_Science2008, Daou_Nature2010, Lawler_Nature2010, Sato_2017, Okamoto_PRB2010,Shenoy_arXiv2016}, so that
an edge-induced $s$-wave pairing component will exhibit slow spatial decay, leading naturally to a gap variation needed to form pseudo-LLs. 
Tuning near such a 
critical point, or using proximity effect coupling to an $s$-wave SC, can tune the decay length and amplitude of the
$s$-wave gap,
thus controlling the pseudo-magnetic field and permitting further experimental tests.

\section{Summary} 

We have proposed inhomogeneously strained nodal SCs as systems to realize pseudo-gauge fields and pseudo-LLs for 
Bogoliubov QPs, and suggested experimental routes and signatures to observe such physics in candidate materials such as the 
cuprate $d$-wave SCs. We note that even accidental SC Dirac nodes will show similar physics.
Further research directions include understanding the impact of such inhomogeneous strains
on the superconducting transition temperature, its interplay with real magnetic fields and vortices,
and extensions to materials like CeCoIn$_5$, iron pnictides, and candidate topological SCs like Sr$_2$RuO$_4$.

{\it Note Added:} After submission of our manuscript, a closely related work appeared
by Emilian Nica and Marcel Franz (arXiv:1709.01158). Our results, where they overlap, are in agreement.

\acknowledgments

This research was funded by the National Science and Engineering Research Council of Canada. AP acknowledges the
support and hospitality of the International Center for Theoretical Sciences (Bangalore) during the completion of this manuscript.

\begin{widetext}

\appendix

\section{Absence of ``scalar gauge potential'' in BdG equation}

Inhomogeneous strain effects also lead to a deformation potential, which in graphene produces a \emph{scalar} gauge potential in addition to the pseudo-vector potential~\cite{Neto2009Electronic, Vozmediano2010Gauge, Lukose_PRL2007, Sanjuan_PRB2014, Castro_2016, Naumis_RPP2017}. 
Here, we argue that no such scalar potential -- which may significantly alter the low-energy LL structure, or even cause its collapse~\cite{Castro_2016} -- can arise in time-reversal symmetric spin-singlet superconducting systems, such as the one we consider. 

The key physical idea is that the BdG Hamiltonian for a singlet SC with time-reversal symmetry only permits 2 of the 4 Pauli matrices -- the corresponding coefficients are in fact the two components of the vector potential identified in the main body of the Letter. Thus, any analog of the `scalar deformation potential' here will necessarily break time-reversal symmetry or lead to singlet-triplet mixing. Such terms will be allowed in a more general setting, for example if spin orbit coupling is present and inversion symmetry or time-reversal symmetry is broken, but not in the cases studied here.

%\subsection*{Technical details}
The Pauli matrix components that can enter the Hamiltonian of Equation~4 of the manuscript are constrained by symmetry. This is most easily seen by considering the BdG Hamiltonian in real space,
\begin{gather}
H_\text{BdG} = \sum_{i, j} \psi^\dagger_i h^{\vphantom\dagger}_{ij} \psi^{\vphantom\dagger}_{j}, \\
h_{ij} = \begin{pmatrix} d^0_{ij} + d^3_{ij} & \Delta_{ij} \\ \Delta_{ji}^* & d^0_{ij} - d^3_{ij} \end{pmatrix},
\end{gather}
where $\psi_{i}^\dagger = ( c^\dagger_{i \uparrow},  c^{\vphantom\dagger}_{i\downarrow} )$ is the Nambu spinor at site $i$, and $d^{0}_{ij}, d^{3}_{ij}, \Delta_{ij}$ are complex numbers, with hermiticity imposing the constraint that $d^0_{ij} = (d^0_{ji})^*$ and $d^3_{ij} = (d^3_{ji})^*$. 
\begin{itemize}
\item Time-reversal symmetry, which sends $c_{i \uparrow} \to c_{i\downarrow}$, $c_{i\downarrow} \to - c_{i\uparrow}$, and complex-conjugates all complex numbers, leads to the additional restrictions (i) $d^0_{ij}=0$ and (ii) $\Delta^{}_{ij} = \Delta_{ji}^*$.
\item Spin rotation symmetry and singlet pairing further imposes the constraints $d^3_{ij} = (d^3_{ij})^*$ and $\Delta^{}_{ij} = \Delta_{ij}^*$.
\end{itemize}
With these ingredients, the Hamiltonian matrix $h_{ij} = d^3_{ij} \sigma^3 + \Delta^{}_{ij} \sigma^1$, where $d^3_{ij}$ and $\Delta^{}_{ij}$ are real numbers. Thus, time-reversal symmetry and spin-rotation symmetry respectively require that the coefficients of $\sigma^0$ (which corresponds to a valley-odd chemical potential) 
and $\sigma^2$ (which corresponds to a complex pairing component) both vanish. 

Such a Hamiltonian captures a BdG SC with arbitrary spatial modulations in hopping and pairing amplitudes, and an appropriate low-energy `Dirac node' expansion recovers Equation~4 of our manuscript, and only permits the two components of the vector potential which we have shown leads to the formation of pseudo-LLs. Any additional `scalar potential' is thus symmetry forbidden. Breaking such symmetries, for instance with a supercurrent that breaks time-reversal symmetry, leads to a Doppler shift for the QPs, which is an
analog of a `scalar potential'.

\section{Dirac BdG solution - gap variations}

Start with the Hamiltonian at node $\ell=1$ for the case discussed in the main text where pseudo-LLs arise from gap variations.
\be
H = \left[\! - i \vf  \sigma^z \partial_x \!+ \! \beta \vd \sigma^x (x \!-\! \frac{k}{\beta}) \! \right]
\ee
Note that $({\rm sgn}\beta~ \sigma^y)$ anticommutes with this Hamiltonian, so that if $\ket{\Phi}$ is an eigenstate of $H$ with energy $E$,
then $({\rm sgn}\beta ~\sigma^y)\ket{\Phi}$ is a solution with energy $-E$. (Here, ${\rm sgn}\beta=\beta/|\beta|$), This is the BdG particle-hole symmetry.
Let us define
\bea
-i\partial_x &=& i \sqrt{\frac{|\beta| \vd}{2\vf}} (a^\dg-a^\pdg)\\
(x-\frac{k}{\beta}) &=& \sqrt{\frac{\vf}{2 |\beta| \vd}} (a^\dg+ a^\pdg)
\eea
so we get
\bea
\!\!\! H = \sqrt{2 |\beta| \vf\vd} \left[ a^\dg \frac{(\sigma^x {\rm sgn}\beta + i \sigma^z)}{2}  + a^\pdg \frac{(\sigma^x {\rm sgn}\beta - i \sigma^z)}{2}  
 \right]
\eea
Define spinors
\be
\ket{\upa} \equiv \frac{1}{\sqrt{2}} \begin{pmatrix} 1 \\ i~{\rm sgn}\beta \end{pmatrix}; \ket{\dna} \equiv  \frac{1}{\sqrt{2}} 
\begin{pmatrix} 1 \\ -i~{\rm sgn}\beta \end{pmatrix}
\ee
Then Hamiltonian is of the Jaynes-Cummings type,
\be
H = \sqrt{2 |\beta| \vf\vd} \left[ i a^\dg {\cal S}^- - i a {\cal S}^+ \right]
\ee
where ${\cal S}^{\pm}$ act as raising/lowering operators on the above spin-$1/2$ states. Let $\ket{n}$ denote harmonic oscillator states (with $n \geq 0$)
centered at $k/\beta$ which are generated by $a,a^\dg$.
Then, we have a zero energy eigenstate 
\be
\ket{\Phi_0} =\ket{0} \ket{\dna}
\ee
and nonzero energy solutions
\bea
\ket{\Phi_{n \pm }} &=& \frac{\ket{n-1} \ket{\upa} \pm i \ket{n} \ket{\dna}}{\sqrt{2}}
\eea
with respective energies $\pm \sqrt{2 |\beta| \vf \vd n}$. More explicitly, the wavefunctions are given by
\bea
\Phi_{k 0}(x) &=& \frac{1}{\sqrt{2}} \varphi_0(x-\frac{k}{\beta}) \begin{pmatrix} 1 \\ -i {\rm sgn}\beta \end{pmatrix} \\
\Phi_{k n\pm} (x) &=& \frac{1}{2} \begin{pmatrix} \varphi_{n-1}(x-\frac{k}{\beta}) \pm i \varphi_n(x-\frac{k}{\beta}) \\ {\rm sgn}\beta (i \varphi_{n-1}(x-\frac{k}{\beta}) \pm \varphi_n(x-\frac{k}{\beta})) \end{pmatrix}
\eea
where $\varphi_n(x)$ is the $n^{\rm th}$ harmonic oscillator ground state.
We can then define quasiparticle operators $\gamma$ for the node pair $\ell=\pm 1$, so that
\bea
\Psi_{1 \alpha}(\br) =\begin{pmatrix} \psi_{1,\alpha}(\br) \\ \epsilon^\pdg_{\alpha\nu} \psi^\dg_{-1,\nu}(\br) \end{pmatrix} = \frac{1}{\sqrt{L}} \sum_k {\rm e}^{i k y} 
\left[ \gamma_{0,\alpha}(k) \Phi_{k 0}(x) + \sum_{n>0} \begin{pmatrix}  \Phi_{k n +}(x) & \Phi_{k n -} (x) \end{pmatrix} 
\begin{pmatrix} \gamma_{n,1,\alpha}(k) \\ \epsilon^\pdg_{\alpha\nu} \gamma^\dg_{n,-1,\nu}(-k) \end{pmatrix} \right]
\eea
In terms of these, the Hamiltonian is given by
\be
H = \sum_{k,\alpha,n>0} \sqrt{2 |\beta| \vf \vd n} \left( \gamma^\dg_{n 1 \alpha}(k) \gamma^\pdg_{n 1 \alpha}(k) + \gamma^\dg_{n 2 \alpha}(k) \gamma^\pdg_{n 2 \alpha}(k)  \right)
\ee

\section{Dirac BdG solution - hopping variations}

Start with the Hamiltonian at node $\ell=2$ for the case discussed in the main text where pseudo-LLs arise from hopping variations. Assume plane waves
along the $x$-direction. Then
\be
H = \beta \vf  \sigma^z (y+\frac{k}{\beta}) \!- \! i \vd \sigma^x \partial_y
\ee
Let us define
\bea
-i\partial_y &=& i \sqrt{\frac{|\beta| \vf}{2\vd}} (a^\dg-a^\pdg)\\
(y+\frac{k}{\beta}) &=& \sqrt{\frac{\vd}{2 |\beta| \vf}} (a^\dg+ a^\pdg)
\eea
so we get
\bea
\!\!\! H = \sqrt{2 |\beta| \vf\vd} \left[ a^\dg \frac{(\sigma^z {\rm sgn}\beta + i \sigma^x)}{2}  + a^\pdg \frac{(\sigma^z {\rm sgn}\beta - i \sigma^x)}{2}  
 \right]
\eea
Define spinors
\be
\ket{\upa} \equiv \frac{1}{\sqrt{2}} \begin{pmatrix} 1 \\ -i~{\rm sgn}\beta \end{pmatrix}; \ket{\dna} \equiv  \frac{1}{\sqrt{2}} 
\begin{pmatrix} 1 \\ i~{\rm sgn}\beta \end{pmatrix}
\ee
Then Hamiltonian is of the Jaynes-Cummings type,
\be
H = \sqrt{2 |\beta| \vf\vd} \left[ i a^\dg {\cal S}^- - i a {\cal S}^+ \right]
\ee
where ${\cal S}^{\pm}$ act as raising/lowering operators on the above spin-$1/2$ states. Let $\ket{n}$ denote harmonic oscillator states (with $n \geq 0$)
centered at $y=-k/\beta$ which are generated by $a,a^\dg$.
Then, we have a zero energy eigenstate 
\be
\ket{\Phi_0} =\ket{0} \ket{\dna}
\ee
and nonzero energy solutions
\bea
\ket{\Phi_{n \pm }} &=& \frac{\ket{n-1} \ket{\upa} \pm i \ket{n} \ket{\dna}}{\sqrt{2}}
\eea
with respective energies $\pm \sqrt{2 |\beta| \vf \vd n}$. More explicitly, the wavefunctions are given by
\bea
\Phi_{k 0}(y) &=& \frac{1}{\sqrt{2}} \varphi_0(y+\frac{k}{\beta}) \begin{pmatrix} 1 \\ i {\rm sgn}\beta \end{pmatrix} \\
\Phi_{k n\pm} (y) &=& \frac{1}{2} \begin{pmatrix} \varphi_{n-1}(y+\frac{k}{\beta}) \pm i \varphi_n(y+\frac{k}{\beta}) \\ -{\rm sgn}\beta 
(i \varphi_{n-1}(y+\frac{k}{\beta}) \pm \varphi_n(y+\frac{k}{\beta})) \end{pmatrix}
\eea
where $\varphi_n(y)$ is the $n^{\rm th}$ harmonic oscillator ground state.

\section{Tunneling density of states (TDOS)}

\subsection{Uniform case}
The superconducting local TDOS for spin-$\alpha$ for a uniform $d$-wave SC is given by 
\be
N_\alpha(\br,\Omega) = \int \frac{d^2\bk}{(2\pi)^2} \left[ u^2_\bk \delta(\Omega-E_\bk) + v^2_\bk \delta(\Omega+E_\bk) \right]
\ee
where $u_\bk^2 = \frac{1}{2} (1+\xi_\bk/E_\bk)$, $v_\bk^2 = \frac{1}{2} (1-\xi_\bk/E_\bk)$, and $E_\bk = \sqrt{\xi^2_\bk + \Delta^2_\bk}$.
We can linearize the dispersion around the 4 nodes (labelled $\ell=\pm 1, \pm 2$), which leads to
\be
N_\alpha(\br,\Omega) = \sum_\ell \int^\Lambda \frac{d^2q}{(2\pi)^2} \frac{1}{2} \left[ (1+ \frac{\vvf^{(\ell)} \cdot \vec{q}}{\cal E_\bq})
\delta (\Omega - {\cal E_\bq}) + (1- \frac{\vvf^{(\ell)} \cdot \vec{q}}{\cal E_\bq})
\delta (\Omega + {\cal E_\bq}) \right]
\ee
where ${\cal E}_\bq = \sqrt{\vf^2 q_\perp^2 + \vd^2 q_\parallel^2}$ and the momentum cutoff $\Lambda$ ensures the
same total number of momentum states. Doing the integral, we find
\be
N_\alpha(\br,\Omega) = 2 \int^\Lambda \frac{dq_\parallel dq_\perp}{(2\pi)^2} \left[
\delta (\Omega -  \sqrt{\vf^2 q_\perp^2 + \vd^2 q_\parallel^2}) +\delta (\Omega + \sqrt{\vf^2 q_\perp^2 + \vd^2 q_\parallel^2}) \right]
\ee
Rescaling $\vf q_\parallel = Q_1$ and $\vd q_\perp = Q_2$, with $Q = \sqrt{Q^2_1+Q^2_2}$, we find 
\be
N_\alpha(\br,\Omega) = 2 \int^\Lambda \frac{dQ}{2\pi \vf \vd} Q \left[
\delta (\Omega -  Q) +\delta (\Omega + Q) \right]
\ee
with an appropriate choice $\Lambda=\sqrt{\pi \vf \vd}$. Of course, this linearized description will break down at a lower energy scale $\sim \vd/a_0$, where $a_0$ is the
lattice spacing.
This yields, for $|\Omega| \lesssim \vd/a_0 \ll \Lambda$,
\be
N(\br,\Omega) = \sum_{\alpha} N_\alpha(\br,\Omega) = \frac{2 |\Omega|}{\pi \vf \vd} 
\ee

\subsection{Pseudo-Landau Level case: Gap variations}

Consider the gap variation example discussed in the main text. Then, fermions at two of the Dirac points only see a phase change from the vector potential, which does not 
change the density of states, leading to a contribution from $\ell=\pm 2$ given by 
\be
N_2(\br,\Omega) = \frac{|\Omega|}{\pi \vf \vd}.
\ee
This is half the total density of states in the uniform case.
The contribution from the other node pair 
$N_1(\br,\Omega)$ is expected to reflect the formation of pseudo-LLs. The Green function for node pair $\ell = \pm 1$ reduces to
\bea
{\cal G}_\alpha^{(\ell=1)}(\br,i\Omega_m) = \frac{1}{2 L} \sum_k \left[ \frac{\varphi^2_0(x-\frac{k}{\beta})}{i\Omega_m-E_0} +  \frac{1}{2} \sum_{n>0} \left(\varphi^2_n(x-\frac{k}{\beta}) 
+ \varphi^2_{n-1}(x-\frac{k}{\beta})\right) \left(\frac{1}{i\Omega_m-E_n} + \frac{1}{i\Omega_m+E_n} \right)\right]
\eea
where $E_0 = 0$. Summing over spins and $\ell=\pm 1$, this leads to 
\bea
N_1(\br,\Omega) = \frac{2}{L} \sum_k \left[ \varphi_0^2(x-\frac{k}{\beta}) \delta(\Omega) + \frac{1}{2} \sum_{n>0}   \left(\varphi^2_n(x-\frac{k}{\beta}) 
+ \varphi^2_{n-1}(x-\frac{k}{\beta})\right) \left( \delta(\Omega-E_n) + \delta(\Omega+E_n) \right) \right]
\eea
Deep in the bulk, $N_1(\br,\Omega)$ will be independent of $\br$, and we can approximate it as 
\be
N_1(\br,\Omega) \approx \frac{|\beta|}{\pi} \left[\delta(\Omega) + \sum_{n > 0} \left(\delta(\Omega-E_n) + \delta(\Omega+E_n) \right) \right]
\ee
which can be recast in the more compact form
\be
N_1(\br,\Omega) \approx \frac{|\beta|}{\pi} \sum_n \delta(\Omega-\lambda_n) 
\ee
where $n=0,\pm 1, \pm 2, \ldots$, with $\lambda_n = \sqrt{2\beta \vf\vd |n|} {\rm sgn}(n)$.
Thus, the total density of states, $N_1(\br,\Omega)+N_2(\br,\Omega)$ will reflect a combination of the  pseudo-LL spectrum as well as the
Dirac density of states of the uniform $d$-wave SC.

\section{Appendix E. \textit{d}-wave SC in a narrow strip}

In this section we study singular contributions to the TDOS which come
from quantization of the quasiparticle momentum transverse to the
strip. Just in this section, we find it convenient to retain the full BdG equation, and linearize around
the Dirac nodes only at the end.
We begin with the BdG Hamiltonian,
\begin{equation}
  \label{eq:H1}
  \hat H(k_L) =   \left( 
    \begin{array}{cc}
      \xi(k_L, -i\partial_w) & \Delta(k_L, -i\partial_w) \\
      \Delta(k_L, -i\partial_w) & -\xi(k_L, -i\partial_w)
    \end{array} \right),
\end{equation}
where $0<w<W$ is the transverse coordinate, and $k_L, k_W$ will denote momenta
along the strip length and strip width ($L$, $W$ directions) respectively. For a $(110)$ edge, we have
$\xi(k_L,-k_W)=\xi(k_L,k_W)$ and
$\Delta(k_L,-k_W)=-\Delta(k_L,k_W)$. We are looking for states which
obey the strip boundary conditions, i.e., eigenfunctions, $\psi(w)$,
of $\hat H$ which have a vanishing charge density at the strip edges,
$\psi^\dagger(0)\tau^z\psi(0) = \psi^\dagger(W)\tau^z\psi(W)=0$.  A
plane wave eigenfunction with positive eigenvalue
$\veps(k_L,k_W)=\sqrt{\xi^2(k_L,k_W)+\Delta^2(k_L,k_W)}$ is given by
\begin{equation}
  \label{eq:phip}
  \phi^+(k_L,k_W;w) = 
  \left(\begin{array}{c}
      u(k_L,k_W) \\ v(k_L,k_W)
    \end{array}\right) e^{ik_W w},
\end{equation}
where
\begin{equation}
  \label{eq:u}
  |u(k_L,k_W)|^2 = \frac{1}{2}\left(1+\frac{\xi(k_L,k_W)}
    {\veps(k_L,k_W)}\right),
\end{equation}
and
\begin{equation}
  \label{eq:v}
  |v(k_L,k_W)|^2 = \frac{1}{2}\left(1-\frac{\xi(k_L,k_W)}
    {\veps(k_L,k_W)}\right).
\end{equation}
Since $\Delta(k_L,k_W)$ is a real function for the $d$-wave SC we are
considering, it is sufficient to take $u(k_L,k_W)>0$ and $({\rm
  sign~}v(k_L,k_W)) = ({\rm sign~}\Delta(k_L,k_W))$, thus,
$u(k_L,-k_W) = u(k_L,k_W)$ and $v(k_L,-k_W) = -v(k_L,k_W)$. A plane
wave eigenfunction with negative energy $-\veps(k_L,k_W)$ is given by
\begin{equation}
  \label{eq:phim}
  \phi^-(k_L,k_W;w) = 
  \left(\begin{array}{c}
      v(k_L,k_W) \\ -u(k_L,k_W)
    \end{array}\right) e^{ik_W w}.
\end{equation}
To construct a state which obeys the boundary conditions, we consider
a superposition of states with opposite $k_W$,
\begin{eqnarray}
  \label{eq:psip}
  \psi^+(k_L,k_W>0,w) & = & \phi^+(k_L,k_W,w)+r(k_L,k_W)\phi^+(k_L,-k_W) 
  \nonumber \\ & = &  \left(\begin{array}{c}
      u(k_L,k_W) \\ v(k_L,k_W)
    \end{array}\right) e^{ik_W w} + r(k_L,k_W)
  \left(\begin{array}{c}
      u(k_L,k_W) \\ -v(k_L,k_W)
    \end{array}\right) e^{-ik_W w}. \nonumber \\
\end{eqnarray}
The charge density for this state is given by (dependence on $k_L$ and
$k_W$ implicit)
\begin{eqnarray}
  \label{eq:rhop}
  \rho^+(w) & = & \psi^{+\dagger}(w)\tau^z\psi^+(w).\
  \nonumber \\ & = & u^2(1 + |r|^2  + 2\Re(re^{-i2k_W w})) 
  - v^2(1 + |r|^2 -2\Re(re^{-i2k_w W})) 
  \nonumber \\  & = & (u^2-v^2)(1+|r|^2)+2\Re(re^{-i2k_W w}),
\end{eqnarray}
where $\Re(z)$ denotes the real part of $z$. Finite size quantization sets as usual $k_W=\pi n /W$,
where $n=0,1,2...$, 
while demanding that $\rho^+$
vanish at the strip edges results in
\begin{equation}
  \label{eq:bcp}
  (u^2-v^2)(1+|r|^2)+2\Re(r) = 0.
\end{equation}
Since, $-1\le u^2-v^2 \le 1$, $r$ is always real. Thus, the
eigenstates are given by
\begin{eqnarray}
  \label{eq:psipn}
  \psi_n^+(k_L;w) & = & \left(\begin{array}{c}
      u_n(k_L) \\ v_n(k_L)
    \end{array}\right) e^{i\pi n w/W} + r_n(k_L)
  \left(\begin{array}{c}
      u_n(k_L) \\ -v_n(k_L)
    \end{array}\right) e^{-i\pi n w/W}.
\end{eqnarray}
Similar states with negative energy are given by
\begin{eqnarray}
  \label{eq:psinn}
  \psi_n^-(k_L;w) & = & \left(\begin{array}{c}
      v_n(k_L) \\ -u_n(k_L)
    \end{array}\right) e^{i\pi n w/W} + r_n(k_L)
  \left(\begin{array}{c}
      -v_n(k_L) \\ -u_n(k_L)
    \end{array}\right) e^{-i\pi n w/W}.
\end{eqnarray}
The TDOS is given by
\begin{equation}
  \label{eq:LDOS}
  N(\Omega,w) =  \frac{1}{2}\int\frac{dk_L}{2\pi}\sum_{n=0}^\infty 
  \sum_{s=\pm} \psi^{s\dagger}_n(k_L,w)(\tau^0+s\tau^z)\psi^s_n(k_L,w)
  \delta(\Omega-s\veps_n(k_L))
\end{equation}
Focusing on positive energies,
\begin{equation}
  \label{eq:LDOSp}
  N(\Omega>0,w) = 2\int\frac{dk_L}{2\pi}\sum_n
  u_n^2(k_L)\left(1+r_n^2(k_L)+2r_n(k_L)\cos\frac{\pi n w}{W}\right)
  \delta(\Omega-\veps_n(k_L))
\end{equation}
The main low energy contributions to the TDOS in a $d$-wave SC come from the
vicinity of the nodes. We are further focusing on the nodes at
$k_L=0$ and $k_W=\pm K_F$, thus, for $k_W>0$, $\xi\simeq
\vf (n\pi/W-K_F)$, $\Delta\simeq v_\Delta k_L$, and
$\veps\simeq\sqrt{\vf^2(n\pi/W-K_F)^2+v_\Delta^2k_L^2}$. Changing
integration variables we have
\begin{eqnarray}
  \label{eq:LDOSp2}
  N(\Omega>0,w) & = & 2\int_{\vf |n\pi/W-K_F|}^\infty\frac{\veps d\veps}{2\pi}\sum_n
  \frac{1}{v_\Delta k_n(\veps)}u_n^2(\veps)\left(1+r_n^2
    +2r_n\cos\frac{\pi n w}{W}\right) \delta(\Omega-\veps) \nonumber \\
  & = & \sum_n\Theta(\Omega-\vf |n\pi/W-K_F|)
  \frac{\Omega}{\pi}\frac{1}{v_\Delta k_n(\Omega)}
  u_n^2(\Omega)\left(1+r_n^2+2r_n\cos\frac{\pi n w}{W}\right),
\end{eqnarray}
where $k_n(\Omega)=\sqrt{\Omega^2-(\vf n\pi/w-K)^2}/v_\Delta$, and
$u_n^2(\Omega) = (1+\vf (n\pi/w-K_F)/\Omega)/2$. Since there are always
values of $w$ for which the term in the above parentheses is
finite, we find that there are contributions at
$\Omega=\vf |n\pi/W-K_F|$ which diverge as
$1/\sqrt{\Omega^2-(\vf n\pi/W-K_F)^2}$.

\section{Appendix F. Pseudo-Landau levels of strained \textit{d}-wave SC in the (1,0)-edged strip geometry}

Numerical diagonalization of the lattice BdG Hamiltonian was also performed for a $(1,0)$-edged strip. Again, the strip's width is $W$, and the transverse direction, along which periodic boundary conditions were used, has length $L \gg W$. Parameters $t$, $t'$, $\bar{n}$, and $\Delta_d$ are taken to be the same as in the $(1,1)$-edged case considered in the main text.

Fig.~\ref{fig:Strip10Spectrum}(a) shows the spectrum of the strip as a function of the momentum $k_L$ along the long direction $L$ in the absence of any imposed spatial variation. The spectrum exhibits the $d$-wave Dirac nodes projected onto the Brillouin zone of the strip. As expected with $(1,0)$ edges, zero-energy ABSs are absent from the spectrum. A circle indicates the near-node region in which we have chosen to plot the spectra of panels (b) and (c).

\begin{figure}[tb]
\centering
\makebox[0.1\textwidth]{\includegraphics[width=0.5\textwidth]{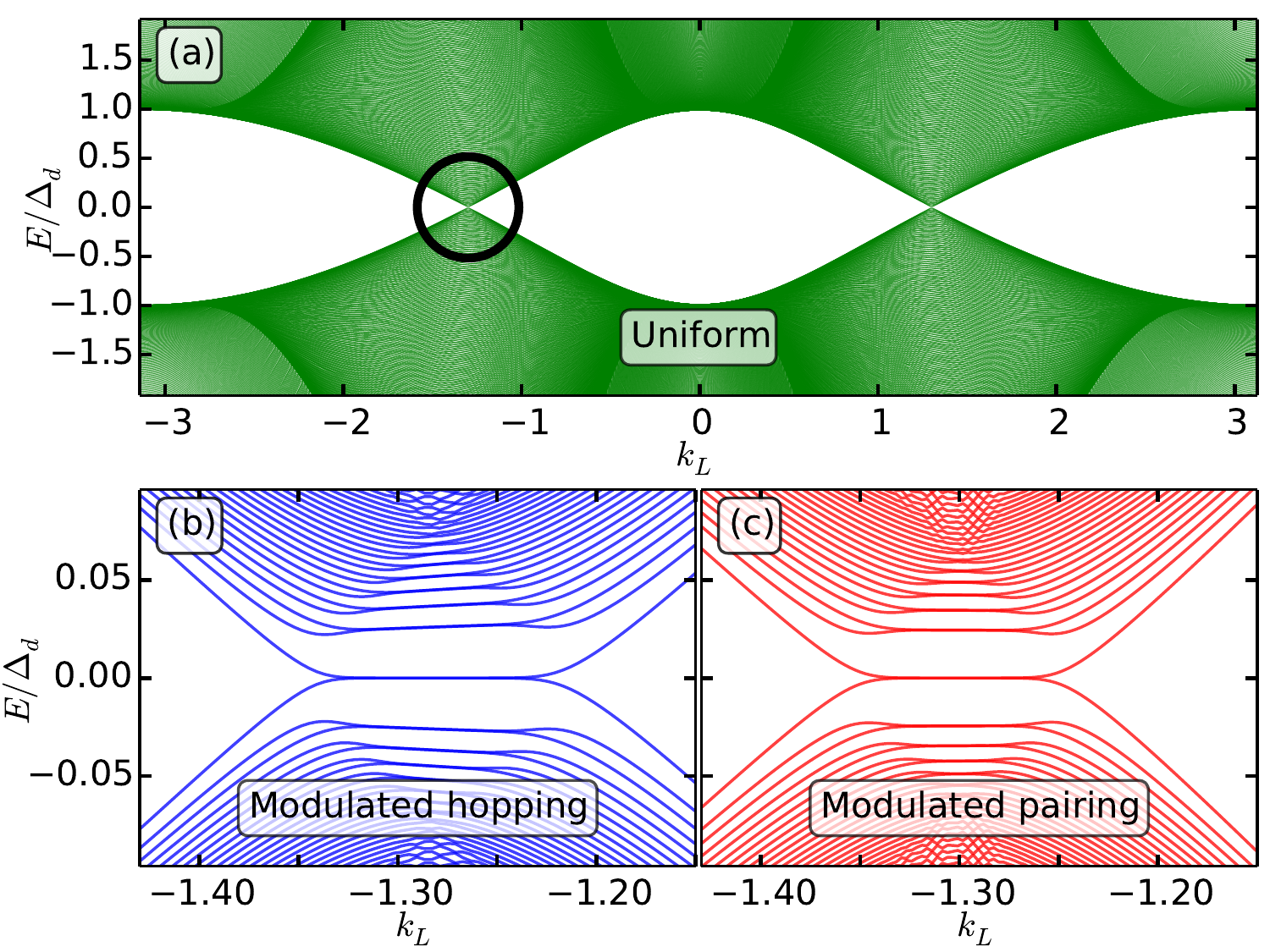}}
\caption{(Color online) 
(a) Spectrum of uniform $d$-wave SC on a $(1,0)$-edged strip versus momentum $k_L$ along the $L$-direction showing Dirac nodes. Note that there are no zero energy ABSs in this geometry. Circle indicates region shown in the next two panels.
(b) Formation of flat pseudo-Landau levels in the low-energy regime due to uniform hopping-amplitude gradient in the $[1,0]$ direction; shown here is the near-node region indicated in (a). 
(c) Similar to (b) but with extended $s$-wave pairing gradient.}
\label{fig:Strip10Spectrum}
\end{figure}

Fig.~\ref{fig:Strip10Spectrum}(b) shows the spectrum in the presence of a nonzero gradient in the hopping amplitude across the strip width (in the $[1,0]$ direction), which leads to a pseudo-LL spectrum at both Dirac nodes; we have chosen $W\!=\! 3000 a_0$ and a maximum change $\delta t \!\sim \! 0.25 t$ at the edge. 
Fig.~\ref{fig:Strip10Spectrum}(c) shows the effect of an extended $s$-wave pairing gradient across the strip width, also leading to pseudo-LL formation at both Dirac nodes. Here, we have chosen $W\!=\! 3000 a_0$ and a maximum $s$-wave gap $\Delta_s \!\sim\! 0.25 \Delta_d$ at the edge. The low energy spectra in Fig.~\ref{fig:Strip10Spectrum}(b) and (c) are in quantitative agreement with our analytical results.

\section{Appendix G. Mean field equations for correlated \textit{d}-wave SC with strain}

We start from the usual $tJ$ model in the main text
\be
H_{tJ} = - g_t \sum_{i,j,\alpha} t_{0,ij} c^\dg_{i\alpha} c^\pdg_{j\alpha} + g_J J \sum_{\la ij \ra} \vec S_i \cdot \vec S_j
\ee
where the {\it bare} nearest neighbor and next-neighbor hoppings are $t_0=1$ and $t'_0=-0.3 t_0$ respectively, 
the antiferromagnetic exchange coupling $J=4 t_0^2/U=0.3 t_0$, and the renormalization factors $g_t=2 p /(1+p)$, $g_J=1$ account for
strong correlation effects in a mean field manner. Note that  $g_t$ is chosen in line with renormalized mean field theory, while we have
set $g_J=1$ similar to what one expects from slave boson mean field theory. At any rate, we should only view this as an effective model
to obtain a variational $d$-wave superconducting ground state, with results which approximately reproduce experimental data.
Doing a full 
Hartree-Fock-Bogoliubov mean field theory of the superexchange term, we arrive at the mean field Hamiltonian
\be
H_{\rm MFT} = \sum_{\bk\alpha} \xi^\pdg_\bk c^\dg_{\bk\alpha} c^\pdg_{\bk\alpha} - 
\sum_\bk \Delta^\pdg_\bk (c^\dg_{\bk\upa} c^\dg_{-\bk\dna} + c^\pdg_{-\bk\dna} c^\pdg_{\bk\upa}),
\ee
where $\xi_\bk = -2 (g_t t_0 + \frac{3}{4} g_J J \chi)  (\cos k_x + \cos k_y) - 4 g_t t_0' \cos k_x \cos k_y$ is set by the effectively renormalized
hoppings (which appear in our BdG calculations in the paper),
$t=(g_t t_0 + \frac{3}{4} g_J J \chi)$ and $t' = g_t t_0'$, while
the pairing gap $\Delta_\bk = \frac{3}{2} g_J J \Delta_0 (\cos k_x - \cos k_y)$.
The mean field equations determining $\chi, \Delta_0$ and the mean electron density $\bar{n} \equiv 1-p$ are given by
\bea
\Delta_0 &=& \frac{1}{2 N} \sum_\bk \frac{\Delta_\bk}{2 E_\bk} (\cos k_x - \cos k_y) \\
\chi &=& \frac{1}{4 N} \sum_\bk (1-\frac{\xi_\bk}{E_\bk}) (\cos k_x + \cos k_y) \\
\bar{n} &=& \frac{1}{N} \sum_\bk (1-\frac{\xi_\bk}{E_\bk}) \equiv 1-p
\eea
where $E_\bk = \sqrt{\xi_\bk^2 + \Delta_\bk^2}$. We solve these equations self-consistently assuming $t_0 \to t_0 (1+\varepsilon)$ and $J \to J (1+2 \varepsilon)$, where
the (small) fractional change $\varepsilon$ in the hopping and exchange interaction is determined by the strain which affects the lattice constant; see main text.
(The factor of $2\varepsilon$ in $J$ reflects its dependence on hopping as $\sim\! t_0^2$.)

We pick the bare hopping $t_0=450$meV, which leads to $J = 135$meV (corresponding to $U/t_0\! \approx\! 13$).
For hole doping $p=0.15$, and for the unstrained case $\varepsilon=0$, we find that the renormalized hoppings satisfy
$t'=-0.25 t$, and
an anti-nodal gap $3 g_J J \Delta_0  \approx 24$meV at $(\pi,0)$. In addition, with the lattice constant 
$a_0=3.85$\AA, we find a nodal Fermi velocity $\vf \approx 1.3$eV-\AA, and a ratio of Fermi velocity to gap velocity $\vf/\vd \approx 20$.
These are in reasonable agreement with results for the optimally doped cuprates. Incorporating $\varepsilon$, and solving the mean field equations, we
find the results for the strain dependence of the hopping and pairing quoted in the main text.

\end{widetext}

%\bibliography{Strain.bib}


\end{document}